\documentclass[aps,floatfix,prd,twocolumn,a4paper,10pt,citeautoscript,
preprintnumbers,superscriptaddress,showpacs]{revtex4}

\usepackage[english]{babel} 	%%% LaTeX for English
\usepackage{amsmath}        	%%% Mathematics AMS package
\usepackage{amssymb}        	%%% More math symbols
\usepackage{bm}				%%% Bold math font
\usepackage{bbm}            	%%% Font for ensembes

\usepackage{graphicx}       	%%% Include graphics
\usepackage{graphics}       	%%% Include graphics
\DeclareGraphicsExtensions{.png,.jpg,.pdf,.eps} %%% Extensions for pdflatex
\usepackage[colorlinks=true, pdfstartview=FitV,linkcolor=blue, citecolor=blue, urlcolor=blue]{hyperref}
\renewcommand{\theequation}{\arabic{equation}}
\newcommand{\Equation}[2]{\begin{equation}\label{#1}#2\end{equation}}
\newcommand{\Align}[2]{\begin{align}\label{#1}#2\end{align}}
\newcommand{\SubAlign}[2]{\begin{subequations}\label{#1}\begin{align}#2\end{align}\end{subequations}}

\newcommand{\bs}{\boldsymbol}
\newcommand{\Figref}[1]{Fig.~\ref{#1}}
\newcommand{\Eqref}[1]{\eqref{#1}}
\newcommand{\groupU}[1]{\mathrm{U}(#1)}  %Unitary group
\newcommand{\Exp}[1]{\text{e}^{#1}}

\renewcommand\Im{\mathrm{Im}}
\newcommand{\Grad}{{\bs\nabla}}

\newcommand{\Curl}{{\bs\nabla}\times}

\newcommand{\A}{{\bs A}}
\newcommand{\B}{{\bs B}}
\newcommand{\D}{{\bs \Pi}}
\newcommand{\E}{{\bs E}}
\newcommand{\F}{\mathcal{F}}

\newcommand{\x}{\bs x}

\newcommand{\Hc}[1]{\mathrm{H}_{c#1}}
\begin{document}
%%%%%%%%%%%%%%%%%%%%%%%%%%%%%%%%%%%%%%%%%%%%%%%%%%%%%%%%%%%%%%%%%%%%%
%%%%%%%%%%%%%%%%%%%%%%%%%%%%%%%%%%%%%%%%%%%%%%%%%%%%%%%%%%%%%%%%%%%%%
%%%% Title informations and authors
\title{Thermoelectric Signatures of Time-Reversal Symmetry Breaking States in 
Multiband Superconductors}
\author{Julien~Garaud}
\affiliation{Department of Theoretical Physics and  Center for Quantum Materials,
KTH-Royal Institute of Technology, Stockholm, SE-10691 Sweden}
\email{garaud.phys@gmail.com}
\author{Mihail~Silaev}
\affiliation{Department of Theoretical Physics and  Center for Quantum Materials,
KTH-Royal Institute of Technology, Stockholm, SE-10691 Sweden}
\author{Egor~Babaev}
\affiliation{Department of Theoretical Physics and  Center for Quantum Materials,
KTH-Royal Institute of Technology, Stockholm, SE-10691 Sweden}
\date{\today}

%%%%%%%%%%%%%%%%%%%%%%%%%%%%%%%%%%%%%%%%%%%%%%%%%%%%%%%%%%%%%%%%%%%%%
%%%% The abstract
\begin{abstract}

We show that superconductors with broken time-reversal symmetry have very specific 
magnetic and electric responses to inhomogeneous heating. A local heating of such 
superconductors induces a magnetic field with a profile that is sensitive to the 
presence of domain walls and crystalline anisotropy of superconducting states. 
A nonstationary heating process produces an electric field and a charge imbalance 
in different bands. These effects can be measured and used to distinguish $s+is$ 
and $s+id$ superconducting states in the candidate materials such as 
Ba$_{1-x}$K$_x$Fe$_2$As$_2$.

\end{abstract}

\maketitle

In many recently discovered superconducting materials, the pairing of electrons 
is supposed to take place in several sheets of a Fermi surface formed by overlapping 
electronic bands \cite{MultibandMgB2,SrRuO,Kamihara.Watanabe.ea:08,Pnictides2,
Pnictides3,Pnictides4}.
Of special interest are the states where the difference of gap's phases in the 
bands is neither $0$ or $\pi$ \cite{Balatsky:00,Lee.Zhang.Wu:09,Zhang2,
StanevTesanovic,Fernandes,Agterberg,Ng,Lin,Johan,Bobkov,Chubukov2,ChubukovMaitiSigrist}. 
Indeed, in addition to the breakdown of usual $\groupU{1}$ gauge symmetry, such 
superconducting states are characterized by an extra broken time-reversal symmetry 
(BTRS) that has numerous interesting physical consequences, many of which are not 
yet explored. Iron-based superconductors \cite{Kamihara.Watanabe.ea:08} are among 
the most commonly accepted candidates for the observation of a BTRS state originating 
from the multiband character of superconductivity and several competing pairing channels.

Experimental data suggest that in the hole-doped 122 compounds Ba$_{1-x}$K$_x$Fe$_2$As$_2$ 
the symmetry of the superconducting state can change depending on the doping level $x$. 
At moderate doping $x\sim 0.4$ various measurements including neutron scattering 
\cite{Christianson.Goremychkin.ea:08}, thermal conductivity \cite{Luo.Tanatar.ea:09}, 
and angle-resolved photoemission spectroscopy (ARPES) \cite{Ding.Richard.ea:08,
Khasanov.Evtushinsky.ea:09,Nakayama.Sato.ea:11} are consistent with the hypothesis 
of the $s_\pm$ state where the superconducting gap changes sign between electron 
and hole pockets.
On the other hand, the symmetry of the superconducting state at heavy doping 
$x\rightarrow 1$ is not so clear regarding the question of whether the $d$ channel 
dominates or if the gap retains $s_\pm$ symmetry changing sign between the inner 
hole bands at the $\Gamma$ point \cite{KorshunovSwave1,KorshunovSwave2}. Indeed, 
there are evidences that $d$-wave pairing channel dominates \cite{Reid.Juneau-Fecteau.ea:12,
ExperimentsDiS1,ExperimentsDiS2,ExperimentsDiS3}, while other ARPES data were 
interpreted in favor of an $s$-wave symmetry \cite{SWaveHoleDoped1,SWaveHoleDoped2}.

%%%%%%%%%%%%%%%%%%%%%%%%%%%%%%%%%%%%%%%%%%%%%%%%%%%%%%%%%%%%%%%%%%%%%%%%%%%%%%%%%%%%%%%%
\begin{figure}[!htb]
\hbox to \linewidth{\hss
\includegraphics[width=.9\linewidth]{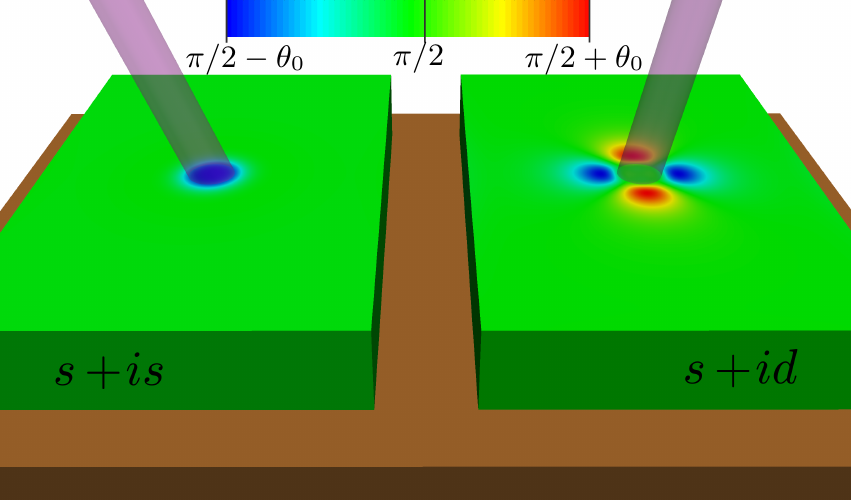}
\hss}
\caption{ %(Color online) --
Variation of the interband phase difference $\theta_{12}$ in BTRS three-band 
superconductors induced by a hot spot created, e.g., by a laser pulse. The phase 
difference variation induced by temperature gradients around hot spots in the 
case of an $s+is$ state (left) preserves $C_4$ symmetry, while it has fourfold 
structure for an $s+id$ state. The value of $\theta_0$ in the $s+is$ case is 
$0.18$ while for $s+id$ it is smaller: $\theta_0=0.05$.  
} \label{Fig:Schematic}
\end{figure}
%%%%%%%%%%%%%%%%%%%%%%%%%%%%%%%%%%%%%%%%%%%%%%%%%%%%%%%%%%%%%%%%%%%%%%%%%%%%%%%%%%%%%%%%

Which of these two possibilities is realized at heavy doping depends on the fine 
balance of the pairing interactions in different channels. However, both cases 
strongly suggest the existence of an intermediate superconducting state that breaks 
time-reversal symmetry at a certain range of the doping level $x$. Two alternative 
scenarios have been considered: namely, $s+id$ and $s+is$ symmetries 
\cite{Lee.Zhang.Wu:09,Zhang2,Chubukov2,SWaveHoleDoped1,SWaveHoleDoped2}.
The $s+id$ state is anisotropic, as it breaks $C_4$ crystalline symmetry,
while the $s+is$ state is qualitatively different as $C_4$ symmetry is
preserved \cite{Chubukov2}. 
Note that the $s+id$ state is qualitatively different from the 
(time-reversal-preserving) $s+d$ state which earlier attracted interest 
in the context of high-temperature cuprate superconductors (see, e.g., 
\cite{Joynt:90,Li.Koltenbah.ea:93,Berlinsky.Fetter.ea:95,Misko.Fomin.ea:00}). 
It also contrasts with the $d+id$ state that violates both parity and 
time-reversal symmetries \cite{Balatsky:00,Franz.Tesanovic:98}. 

To this day, no experimental proof of $s+is$ or $s+id$ BTRS states 
has been reported. Indeed, probing the relative phases between components 
of the order parameter in different bands is a challenging task.
For example, the $s+is$ state does not break point group symmetries and 
therefore it is not associated with intrinsic Cooper pair angular momentum. 
Hence, it cannot produce a local magnetic field and is invisible for conventional 
methods like muon spin relaxation and polar Kerr effect measurements that were 
used to search for BTRS $p+ip$ superconducting state in, e.g., the Sr$_2$RuO$_4$ 
compound \cite{Mackenzie.Maeno:03}. 
Proposals for indirect observation of a BTRS signature in pnictides, with various 
limitations, have been recently voiced. These include, for example, investigation 
of the spectrum of collective modes which includes massless \cite{Lin} and 
mixed phase-density \cite{Johan,Stanev,Chubukov2,Benfatto} excitations.
It was also proposed to consider exotic topological excitations in the form of 
skyrmions and domain walls \cite{Garaud.Carlstrom.ea:11,Garaud.Carlstrom.ea:13,
Garaud.Babaev:14}, an unconventional vortex viscosity mechanism \cite{Silaevvisc}, 
vortex clustering \cite{Johan}, and exotic reentrant and precursor phases induced 
by fluctuations \cite{bojesen2013time,bojesen2014phase,carlstrom2014spontaneous,
hinojosa2014time}.
Spontaneous currents were predicted to exist near impurities in anisotropic 
superconducting states \cite{Lee.Zhang.Wu:09,ChubukovMaitiSigrist} or in samples 
subjected to strain \cite{ChubukovMaitiSigrist}. The latter proposal actually 
involves symmetry change of $s+is$ states and relies on the presence of disorder, 
which usually has an uncontrollable distribution.

In this Letter, we discuss an experimental set-up based on a local heating, that 
allows the direct observation of BTRS states in a controllable way. This is illustrated 
in \Figref{Fig:Schematic}, where local heating induces a local variation of relative 
phases that are further shown to yield electromagnetic excitations. The key idea 
is based on the recent proposal of an unconventional thermoelectric effect in 
BTRS multiband superconductors \cite{Silaev.Garaud.ea:15}. There, a temperature 
gradient generates phase gradients of condensate components, due to the generically 
temperature-dependent interband phase differences $\theta_{ki}(T):=\theta_{k}
-\theta_{i}$ (where $k,i$ are band indices). It results that the local heating 
generates spontaneous magnetic fields and charge imbalance distributions. These 
thermoelectric responses are drastically different from their counterparts in 
conventional superconductors \cite{Maniv.Polturak.ea:05}. As discussed below, 
the fields created by local heating have opposite directions in two degenerate 
superconducting states [i.e. $s+is(d)$ and $s-is(d)$ ones]. They are measurable 
by conventional techniques (e.g. by SQUID) and therefore scans of the surface 
can be used to diagnose the structure of order parameter and interband phase 
differences, to detect pinned domain walls or broken crystalline symmetry 
states in either $s+is$ or $s+id$ superconductors.

%%%%%%%%%%%%%%%%%%%%%%%%%%%%%%%%%%%%

We consider a minimal three-band microscopic model which has been suggested to
describe the BTRS superconducting state in the hole-doped 122 iron-pnictide compounds
\cite{StanevTesanovic,Benfatto,Chubukov2} with three distinct superconducting 
gaps $\Delta_{1,2,3}$ in different bands. The pairing which leads to the BTRS 
state is dominated by the competition of different interband repulsion channels 
described by the following coupling matrix: 
\begin{equation}\label{Eq:Model3BandText}
 \hat g = - \nu_0  \left(%
 \begin{array}{ccc}
 0        & \eta    & \lambda   \\
 \eta     & 0       & \lambda   \\
 \lambda  & \lambda & 0         \\
 \end{array}  \right) \,.
\end{equation}
Here, we assume for simplicity that the density of states $\nu_0$ is the same 
in all superconducting bands. This model has been suggested \cite{StanevTesanovic,
Benfatto,Chubukov2} in order to describe transitions between $s/s_{\pm}$ and 
$s+is$ states when tuning parameters $\eta$,$\lambda$ and temperature. The 
dimensionless coefficients $\eta$ and $\lambda$ describe different pairing channels, 
whether it is an $s+is$ or an $s+id$ state. In the former case $\Delta_{1,2}$ 
correspond to the gaps at hole Fermi surfaces and $\Delta_3$ is the gap at the electron 
pockets, so that $u_{hh}= \nu_0\eta$ and $u_{eh} = \nu_0\lambda$ are, respectively, 
the hole-hole and electron-hole interactions \cite{Benfatto,Chubukov2}. The same model 
\Eqref{Eq:Model3BandText} can be used to describe the $s+id$ states but there, 
$\Delta_{1,2}$ describe gaps in electron pockets and $\Delta_3$ is the gap at the 
hole Fermi surface, so that $u_{eh} = \nu_0\lambda$ and $u_{ee} = \nu_0\eta$ are 
electron-hole and electron-electron interactions respectively. 

%%%%%%%%%%%%%%%%%%%%%%%%%%%%%%%%%%%%
To study magnetic and electric responses of both $s+is$ and $s+id$ states, 
we use a Time-Dependent Ginzburg-Landau (TDGL) approach \cite{KopninBook, Dorsey} 
generalized to a multiband case (see derivation in Appendix \ref{Appendix:GL} and  
\ref{Appendix:TDGL}). The dimensionless TDGL equations read (see details in 
\footnote{
The lengths are normalized by  $\tilde{\xi}_0 = \hbar \bar{v}_F/T_c$, 
where $\bar{v}_F$ is the average value of Fermi velocity, magnetic field 
by $B_0 = T_c\sqrt{\nu_0/\rho}$, where $T_c$ is in energy units and 
$\rho = 7\zeta(3)/(8\pi^2) \approx 0.1$. The magnetic field scale $B_0$ 
is of the order the thermodynamic critical field at low temperatures 
\cite{Saint-James.Thomas.ea}. In such units the electron charge is replaced 
by an effective coupling constant $\tilde{e} = \pi B_0\xi_0^2/\Phi_0$. 
}\nocite{Saint-James.Thomas.ea}):
\begin{equation}\label{Eq:TDGL}
 (\partial_t+2i\tilde{e} \varphi)\psi_k=
 -\frac{\delta\F}{\delta\psi_k^*}
 \,,~~~ \Curl\B -\sigma_n \E={\bm j}_s \,,
\end{equation}
where  $\varphi$ is the electrostatic potential, $\sigma_n$ is the normal state 
conductivity, and ${\bm j}_s = - \delta\F/\delta\A$ is the superconducting current. 
Near the critical temperature the energy relaxation is determined by the phonon 
scattering which yields the relaxation time scale 
$t_0 = \pi\hbar /(8T_c) \sim 1~\mathrm{ps}$ provided $T_c \sim 1~\mathrm{meV}$, 
which is about $10~\mathrm{K}$.   

Note that multiband superconductors are described by several components 
$\psi_{k}$ which do not necessarily coincide with the gap functions $\Delta_i$ 
in different bands (see, e.g., \cite{Benfatto,Chubukov2}). For example, since
the coupling matrix \Eqref{Eq:Model3BandText} has only two positive eigenvalues, 
the relevant Ginzburg-Landau (GL) theory reduces to a two-component one (see 
details in Appendix \ref{Appendix:GL}), which in dimensionless units reads as
%%%%%%%%%%%%%%%%%%%%%%%%%%%%%%%%
\begin{subequations}\label{Eq:FreeEnergyText}\begin{align}
 \F &=\frac{\B^2}{8\pi}+\sum_{j=1}^2\left(
    k_j\left|\D\psi_j \right|^2
    +\alpha_j|\psi_j|^2+\frac{\beta_j}{2}|\psi_j|^4\right )
    \label{Eq:FreeEnergy:Self}  \\
   &+k_{12,a}(\Pi^*_a\psi_1^*\Pi_a\psi_2+c.c. )
    \label{Eq:FreeEnergy:MixedText} \\
   &+\gamma|\psi_1|^2|\psi_2|^2
    +\frac{\delta}{2}\big(\psi_1^{*2}\psi_2^2+c.c.\big)
    \label{Eq:FreeEnergy:InteractionText}   \,,
\end{align}\end{subequations}  %%$\D=\Grad-2\pi i\A/\Phi_0$.
with $\D=\Grad-2i\tilde{e}\A$. The components $\psi_{1,2}$ are determined 
by a superposition of the different gap functions $\Delta_i$. All coefficients 
of the model \Eqref{Eq:FreeEnergyText} are consistently determined from the 
microscopic coupling matrix (see details in Appendix \ref{Appendix:GL}), 
and the temperature dependence is given by the coefficients:
\begin{subequations}\label{Eq:alpha}\begin{align}
 \alpha_1 &= -2(G_0-G_1+\tau ) \\
 \alpha_2 &= -(2x^2+1)(G_0-G_2+\tau )
\end{align}\end{subequations}
where $\tau=(1-T/T_c)$, $x= (\eta- \sqrt{\eta^2+8\lambda^2})/(4\lambda)$,
$G_{1,2}$ are the positive eigenvalues of the inverse coupling matrix 
$\nu_0 \hat g^{-1}$ and $G_0=\mathrm{min(G_1,G_2)}$. 
The general GL functional \Eqref{Eq:FreeEnergyText} derived from 
the three-band microscopic model has $\delta>0$. Hence it favors BTRS with 
$\pm \pi/2$ phase differences between $\psi_1$ and $\psi_2$ order parameter 
components, which describes both the $s+is$ and $s+id$ states depending on the 
structure of mixed gradient terms \Eqref{Eq:FreeEnergy:MixedText}. They are 
$k_{12,x}=k_{12,y}$ for the $s+is$ state and $k_{12,x}=-k_{12,y}$ for the 
crystalline $C_4$-symmetry breaking $s+id$ state. 

As a consequence of the discrete degeneracy due to BTRS, the model 
\Eqref{Eq:FreeEnergy:InteractionText} allows domain walls (DW) interpolating 
between regions with different relative phases. The direct observation of 
DWs in $s+is(d)$ states is challenging. Unlike DWs in $p+ip$ superconductors, 
they do not generate a spontaneous magnetic field. However, by our general argument 
below, DWs should provide a controlled magnetic response in the presence of 
relative-density perturbations that can be induced by a local heating.

To investigate the response to spatial modulations of the components of the 
order parameter induced by a local heat source, the fields $\psi_{1,2}$ 
and $\A$ are discretized using a finite-element framework \cite{Hecht:12} 
(see discussion of numerical methods in Appendix \ref{Appendix:Numerics}).
To model the local heating, the temperature profile is found by solving the 
(stationary) heat equation for a heat source at temperature $T_s$, while 
boundaries are kept at $T_0=0.7T_c$. Once the temperature profile is found, 
the coefficient $\alpha_k$ in \Eqref{Eq:FreeEnergyText} varies in space and 
the TDGL equations \Eqref{Eq:TDGL} are evolved for $\Delta t=80$ (in units 
$t_0$ defined above). The temperature of the heat source is then modified 
to $T^\prime_s$, and the TDGL equations are further evolved for the new 
temperature profile for a period $\Delta t$. The temperature of the source 
is initially set to $T_0$, sequentially ramped up to $0.95T_c$, and then ramped 
down back to $T_0$. 
In our simulations, we chose the dimensionless conductivity $\sigma_n=0.1$ 
and the coupling constant $\tilde{e}=0.113$. The coefficients in GL functional 
\Eqref{Eq:FreeEnergyText} are determined using the microscopic coupling matrix 
\Eqref{Eq:Model3BandText} with coefficients $\eta=5$ and $\lambda=4.5$
\footnote{ 
The consistently obtained values are $\beta_1=2$, $\beta_2=1.108$, $\delta=0.465$ 
and $\gamma=0.929$. The coefficients of the kinetic terms are $k_1=0.55$, 
$k_2=0.375$ and $k_{12,x}=0.217$, while $k_{12,y}=k_{12,x}$ for $s+is$ states 
and $k_{12,y}=-k_{12,x}$ for $s+id$. See detailed derivation in Appendix 
\ref{Appendix:GL} and \ref{Appendix:TDGL}.
}.

%%%%%%%%%%%%%%%%%%%%%%%%%%%%%%%%%%%%%%%%%%%%
\begin{figure}[!tb]
\hbox to \linewidth{\hss
\includegraphics[width=\linewidth]{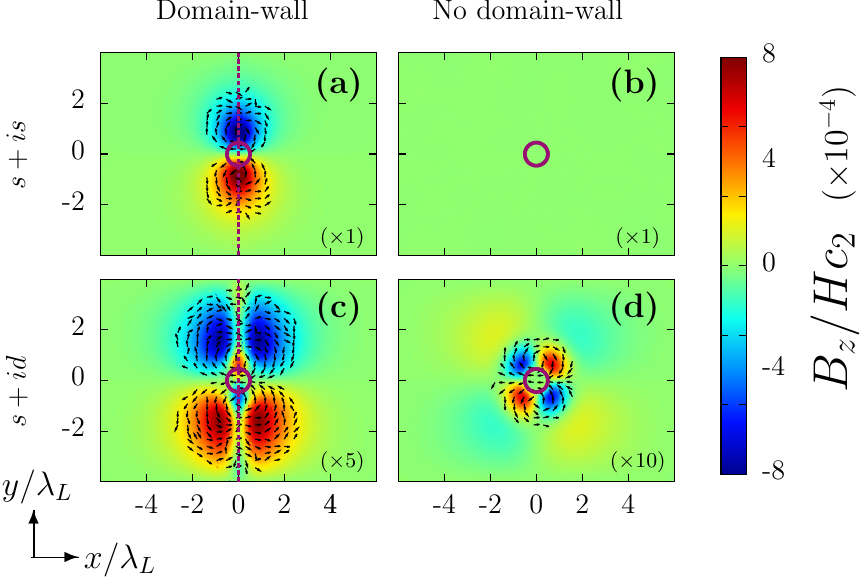}
\hss} \caption{ %(Color online) --
The magnetic response  that originates in local heating of the sample. Panels 
(a) and (c), respectively, show the response when a local hot spot heats an area of 
a sample which contains a domain wall. Here, we display two cases of the domain walls 
for two BTRS states: $s+is$ and $s+id$. Panels (b) and (d) show the response to 
the local hot spot in the case of the homogeneous BTRS state for the $s+is$ and $s+id$ 
superconducting state.
The color plot shows the magnitude of the out-of-plane induced magnetic field 
$B_z$ (magnitudes differ in different panels), while arrows indicate the 
orientation of supercurrents.
The dotted line indicates the presence of the domain wall and the inhomogeneous 
temperature profile is induced by the ringlike heat source shown by the small 
circles. Length scales are given in terms of London penetration length and 
calculated values of the GL parameters are given in \cite{Note1,Note2} and 
Appendix \ref{Appendix:GL}-\ref{Appendix:Numerics}. 
} \label{Fig:Magnetic}
\end{figure}
%%%%%%%%%%%%%%%%%%%%%%%%%%%%%%%%%%%%%%%%%%%%
As shown in \Figref{Fig:Magnetic}, according to the simulations, when the 
source heats up a domain wall, it induces a multipolar magnetic field with 
zero net flux. In the case of a superconductor with $s+is$ symmetry, it shows 
a dipolar structure, while it is differently distributed for $s+id$. On the 
other hand, when the heat source is focused on the uniform $s+is$ state it 
shows no magnetic response, while a fourfold magnetic field is induced in 
the $s+id$ case as a result of the explicit breakdown of the $C_4$ symmetry.
Here, spatial variations are normalized to the penetration depth $\lambda_L$ 
and the amplitudes of the induced magnetic field to the second critical field $\Hc{2}$ 
defined by the GL functional \Eqref{Eq:FreeEnergyText}. Provided the typical 
values of $\Hc{2}\sim 10$T in 122 pnictides \cite{Stewart:11}, one can see 
that the magnetic response can be detected with high accuracy by conventional 
local probes of static magnetic field such as scanning SQUID or Hall probe 
microscopy.

The physical origin of the spontaneous magnetic response follows from that the total 
current is the sum of partial currents in each of the $N$ bands ${\bm j}=\sum_{k=1}^N{\bm j}_k$
and therefore can generated by the gradients of relative phases \cite{Silaev.Garaud.ea:15}. 
Since ${\bm j}_k=(\Grad\theta_{k}-2\pi\A/\Phi_0)c\Phi_0/(8\pi^2\lambda_{k}^2)$
the London expression for the magnetic field in multiband superconductors is modified 
as follows 
\begin{equation} \label{Eq:MagneticField}
 \B = - \frac{4\pi}{c}\Curl \left( \lambda^{2}_L {\bm j} \right) +
 \frac{\Phi_0}{2\pi N}\sum_{k>i}\Curl\left( \gamma_{ki} \Grad\theta_{ki}\right)
\end{equation}
where  $\lambda_{k}$ are coefficients characterizing contribution of each band to 
the Meissner screening, $\lambda_L = 1/\sqrt{\sum_k \lambda^{-2}_{k}}$ is the London 
penetration depth and 
$\gamma_{ki}({\bm r })= \lambda_L^2(\lambda_k^{-2}({\bf r })-\lambda_i^{-2}({\bf r }))$. 
 In contrast to London's magnetostatics, Eq.~\Eqref{Eq:MagneticField} shows that 
the magnetic field features an additional contribution when relative density 
gradients $ \Grad\gamma_{ki} (\bm{r})$ are noncollinear with that of relative-phase 
gradients $\Grad\theta_{ki}$. Such gradients generically appear in BTRS states if a 
domain wall-containing superconductor is exposed to a local heat source.
The second term in \Eqref{Eq:MagneticField} can be nonzero even in the absence 
of domain walls due to direction dependent tensor coefficients 
$\hat \gamma_{ki} ({\bf r})$ in anisotropic $s+id$ states.

Domain walls can be created by quenching the sample and stabilized by 
pinning or artificial geometric barriers \cite{Curran.Bending.ea:14,
Garaud.Babaev:14}. Yet it is also important to obtain the evidence 
of isotropic $s+is$ states for homogeneous superconducting states. 
Below, we show that this can be done by considering the nonequilibrium 
electric responses generated by nonstationary heating when the local 
temperature evolves recovering from the initial hot spot created, e.g., 
by a laser pulse \cite{Maniv.Polturak.ea:05}.
An unusual electric response can be seen when combining
Eq.\Eqref{Eq:MagneticField} to Faraday's law. The electric field
$\E=-c^{-1}\partial_t{\A}-\Grad\varphi$ can be rewritten as
\begin{equation}  \label{Eq:ElectricField}
 {\bm E} = \frac{4\pi}{c^2}\frac{\partial}{\partial t} \left( \lambda_L^2 {\bm j} \right)
 -\frac{\Phi_0}{2\pi N c}\sum_{k>i} \frac{\partial}{\partial t}
  \left( \gamma_{ki} \Grad \theta_{ki}\right)  - \Grad\Phi\,.
\end{equation}
Here, $\Phi=  \sum_k (\varphi + \hbar\dot{\theta}_{k}/2e) /N$ is a 
gauge invariant potential field, determined by the sum of chemical
potential differences between quasiparticles $\mu_q= e\varphi$ and
condensates in each band $\mu^{(k)}_{p} = - \hbar\dot{\theta}_{k}/2$.
Each of the partial potential differences $\Phi^{(k)} = [\mu_{q}
-\mu^{(k)}_{p}]/e$ is proportional to charge imbalance in the $k$-th band
$Q^*_k = 2e^2\nu_0\Phi^{(k)}$ %where $\nu_0$ is the density of states 
\cite{Rieger.Scalapino.ea:71,Tinkham.Clarke:72,Kadin.Smith.ea:80}.
%.
In multicomponent systems the charge imbalance can generated by 
variations of interband phase differences in space and time. The physics 
behind this process is a nonequilibrium redistribution of Cooper pairs 
between different bands which initially creates partial charge imbalances 
$Q^*_k$. This mechanism leads to the unconventional electric response of 
BTRS superconductors to a nonstationary local heating. It can be measured 
with potential probe techniques that were employed to study the imbalance 
between quasiparticles and condensate subsystems in conventional 
superconductors \cite{Clarke:72,Yu.Mercereau:75}.

%%%%%%%%%%%%%%%%%%%%%%%%%%%%%%%%%%%%%%%%%%%%%%%%%%%%%%%%%%%%%%%%%%%%%
\begin{figure}[!tb]
\hbox to \linewidth{\hss
\includegraphics[width=\linewidth]{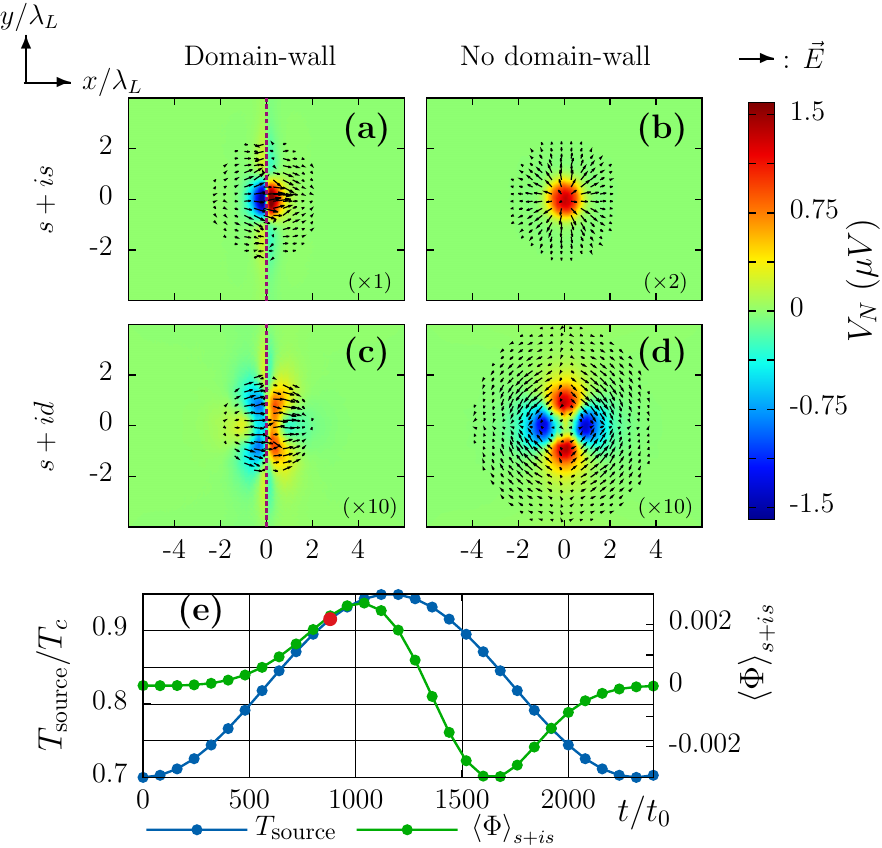}
\hss}
\caption{ %(Color online) --
Electric response due to the nonstationary heating of the sample. Panel (e) 
shows the time evolution of the source's temperature. Panels (a)-(d) correspond 
to the cases in \Figref{Fig:Magnetic} and the color plot shows the voltage $V_N$ 
generated by the charge imbalance that can be picked by a normal detector, 
while the arrows correspond to the electric field. Panel (e) also displays $\Phi$ 
integrated over the whole sample, of an $s+is$ superconductor without DW (b). 
The red dot on panel (e) denotes the ``position'' of panels (a)-(d) in the time 
series.
} \label{Fig:Imbalance}
\end{figure}
%%%%%%%%%%%%%%%%%%%%%%%%%%%%%%%%%%%%%%%%%%%%%%%%%%%%%%%%%%%%%%%%%%%%%

\Figref{Fig:Imbalance} shows such an electric response to a nonstationary
heating of the superconducting sample. This multicomponent electrodynamic
phenomenon can be used to detect BTRS states through charge imbalance
generation in response to nonstationary heating.
As shown in \Figref{Fig:Imbalance}, the charge imbalance shows nontrivial
pattern that is different for $s+is$ and $s+id$ states. The total charge
imbalance $\langle\Phi\rangle$ in uniform $s+id$ is zero as a result of
the fourfold symmetric structure due to broken $C_4$ symmetry. On the other 
hand, as shown in \Figref{Fig:Imbalance}(e), $\langle\Phi\rangle\neq0$ in 
uniform $s+is$. Note that the total imbalance in the case of domain walls 
\Figref{Fig:Imbalance}(a) and \Figref{Fig:Imbalance}(c) is zero because the 
heat source is centered at the domain wall. A shifted source from the DW center 
will not be symmetric and thus its average will not vanish.

The generated charge imbalance can be measured with the standard technique 
using the normal metal and superconducting potential probes 
\cite{Clarke:72,Yu.Mercereau:75}. The magnitude of voltage $V_N$ induced 
in the normal detector is related to the dimensionless signal shown in 
\Figref{Fig:Imbalance} as $V_N = h \Phi / (t_0 e)$, where $h/(t_0 e)\approx4\,\mathrm{mV}$. 
The overall magnitude of the voltage signal is thus expected at the order 
of the $\mu V$. 
The electric field and charge imbalance depend on the dynamics of the temperature 
profile variation, while $\B$ depends on the temperature profile itself. 
As a result, the sign of the induced electric field and charge imbalance 
changes when ramping down the heat-source temperature, while it does not for 
the magnetic field (see animations \cite{Supplementary} and their description 
in Appendix \ref{Appendix:Movie}).
Note that similarly to the spontaneous magnetic field, the induced charge 
imbalances are sensitive to BTRS: degenerate $s+is$ and $s-is$ states 
produce opposite electric fields and charge imbalances in response to the 
same heating protocol. It allows to discriminate between the usual thermoelectric 
occurring in conventional superconductors and the unconventional one being a 
specific signature of BTRS states.

%%%%%%%%%%%%%%%%%%%%%%%%%%%%%%%%%%%%%%%%%%%%%%%%%%%%%%%%%%%%%%%%%%%%%
%%%%%%%%%%%%%%%%%%%%%%%%%%%%%%%%%%%%%%%%%%%%%%%%%%%%%%%%%%%%%%%%%%%%%
%%% CONCLUSION

To conclude, we demonstrated possible direct manifestations of BTRS states 
in experimentally observable electric and magnetic responses to nonuniform 
and nonstationary heating. The signs of the generically induced magnetic 
field and charge imbalance distributions are opposite in degenerate BTRS 
states [i.e. in $s+is(d)$ and $s-is(d)$]. 
These specific thermoelectric behaviors were also shown to reveal the presence 
of domain walls between $s+is(d)/s-is(d)$ states. Moreover, the demonstrated 
crucial dependence of thermomagnetic and charge imbalance responses on crystalline 
anisotropy provides an experimental tool to distinguish between isotropic $s+is$ 
and $C_4$ symmetry-breaking $s+id$ states that are particularly interesting 
for pnictides.

%%%%%%%%%%%%%%%%%%%%%%%%%%%%%%%%%%%%%%%%%%%%%%%%%%%%%%%%%%%%%%%%%%%%%
\begin{acknowledgments}
The work was supported by the Swedish Research Council Grants
No. 642-2013-7837.%and No. 325-2009-7664.
The computations were performed on resources provided by the
Swedish National Infrastructure for Computing (SNIC) at the National
Supercomputer Center at Link\"oping, Sweden.
\end{acknowledgments}

%%%%%%%%%%%%%%%%%%%%%%%%%%%%%%%%%%%%%%%%%%%%%%%%%%%%%%%%%%%%%%%%%%%%
\appendix
\setcounter{section}{0}
\setcounter{paragraph}{0}
\setcounter{equation}{0}
\renewcommand{\theequation}{\Alph{section}.\arabic{equation}}
%%%%%%%%%%%%%%%%%%%%%%%%%%%%%%%%%%%%%%%%%%%%%%%%%%%%%%%%%%%%%%%%%%%%%
%%%%%%%%%%%%%%%%%%%%%%%%%%%%%%%%%%%%%%%%%%%%%%%%%%%%%%%%%%%%%%%%%%%%%

\section{Ginzburg-Landau expansion for the \texorpdfstring{$s+is$}{sis}
and \texorpdfstring{$s+id$}{sid} superconductors} 
\label{Appendix:GL}

\begin{figure}[!htb]
\hbox to \linewidth{\hss
\includegraphics[width=\linewidth]{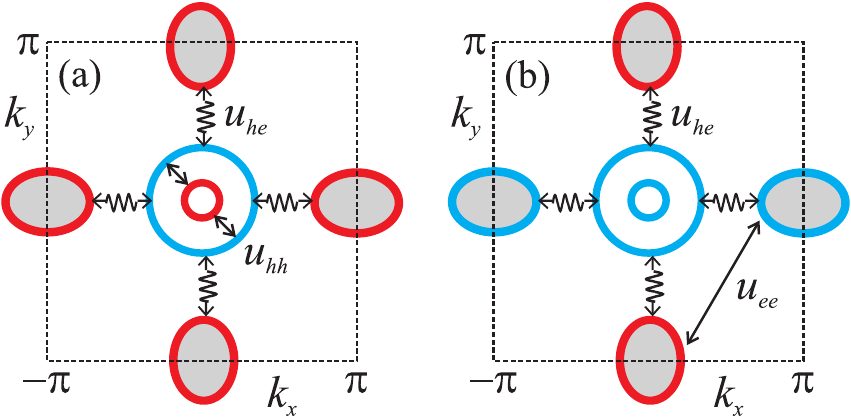}
\hss}
\caption{
(Color online) --
Schematic view of the band structure in
hole-doped iron pnictide compound Ba$_{1-x}$K$_x$Fe$_2$As$_2$. It
consists of two hole pockets at $\Gamma$ shown by open circles and
two electron pockets at $(0; \pi)$ and $(\pi; 0)$ shown by filled
ellipses. In panel (a), the $s+is$ state is favored by superconducting
coupling dominated by the interband repulsion between electron and hole
Fermi surfaces $u_{he}$, as well as between the two hole pockets $u_{hh}$.
In panel (b), there is a possibility of $s+id$ state due to the repulsion
between electron and hole Fermi surfaces $u_{he}$ as well as between
electron pockets $u_{ee}$.
}
\label{Fig:BZ}
\end{figure}

We consider here two alternative patterns of superconducting
coupling which both result on BTRS state but will be shown to
yield qualitatively different physical properties. In the first
scheme shown in \Figref{Fig:BZ}(a) the dominating pairing channels
are the interband repulsion between electron and hole bands,
as well as between two hole pockets at $\Gamma$. There, the order
parameter is the same in both electron pockets so that the
crystalline $C_4$ symmetry is not broken and thus corresponds
to an $s$ state. Instead the second alternative shown in
\Figref{Fig:BZ}(b) is that when the strongest interactions are
the repulsions between hole and electron bands and between two
electron pockets. Such interaction favours order parameter sign
change between electron pockets resulting in a $C_4$ symmetry
breaking $d$-wave state.

To derive the Ginzburg-Landau expansion that is used in our
simulations, we consider the microscopic model of clean
superconductor with three overlapping bands at the Fermi level.
Within quasiclassical approximation the band parameters
characterizing the two different cylindrical sheets of the Fermi
surface are the Fermi velocities $v^{(j)}_{F}$  and the partial
densities of states (DOS) $\nu_j$, labelled by the band indices
$j=1,2,3$. To describe the two possible alternatives of BTRS
states, namely $s+is$ and $s+id$ symmetries, we consider two
three-band models schematically shown in \Figref{Fig:BZ}.
The Eilenberger equations for quasiclassical propagators take 
the form
\begin{align}\label{Eq:EilenbergerF}
 &\hbar{\bm v}^{(j)}_{F}{\bm \Pi} f_j +
 2\omega_n f_j - 2 \Delta_j g_j=0, \\ \nonumber
 &\hbar{\bm v}^{(j)}_{F}{\bm \Pi}^* f^+_j -
 2\omega_n f^+_j + 2\Delta^*_j g_j=0 \,,
\end{align}
where ${\bm \Pi}=\Grad  - 2\pi i \A/\Phi_0$, $\A$ is the vector
potential, ${\bm v}^{(j)}_{F}$ is the Fermi velocity. The
quasiclassical Green's functions in each band obey normalization
condition $g_j^2+f_jf_j^+=1$. The self-consistency equations for
the gaps and electric current are
%
%
%
%%%%%%%%%%%%%% FIGURE %%%%%%%%%%%%%%%%%%
\begin{eqnarray}\label{Eq:SelfConsistentGap}
 \Delta_i ({\bm p},{\bm r})= 2\pi T \sum_{n,{\bm p^\prime},j} \lambda_{ij}({\bm p},{\bm p^\prime})
  f_j ({\bm p},{\bm r}, \omega_n) \\
  \label{Eq:SelfConsistentCurrent}
 {\bm j} ({\bm r})=  2\pi e T \nu \sum_{n, {\bm p},j}
  {\bm v}_F^{(j)} \Im\; g_j ({\bm p},{\bm r}, \omega_n)
\end{eqnarray}
where $g_j=\mathrm{sign} (\omega_n)\sqrt{1- f_jf^+_j}$ and $\nu$ is the 
density of states, ${\bm p}$ parameters run over the corresponding Fermi 
surfaces and $\lambda_{ij}$ is the coupling potential matrix. For simplicity 
we will consider further isotropic pairing states so that 
$\lambda_{ij}({\bm p}, {\bm p^\prime} )= const$ on each of the Fermi surfaces.
However in electron pockets we keep the anisotropy of Fermi velocities in 
Eq.(\ref{Eq:EilenbergerF}). We neglect the anisotropy of hole bands which 
is a well-justified assumption \cite{VorontsovVekhter}.

The derivation of the GL functional from the microscopic equations
formally follows the standard scheme.  First we find the solutions
of Eqs.(\ref{Eq:EilenbergerF}) in the form of the expansion by
powers of the gap functions amplitudes $\Delta_{j}$ and their
gradients:
 \begin{eqnarray}\label{Eq:GLexpansion}
 & f_j({\bm p},{\bm r}, \omega_n) = \\ \nonumber
 & \frac{\Delta_j}{\omega_n}-\frac{|\Delta_j|^2\Delta_j}{2\omega_n^3}-
 \frac{ \hbar( {\bm v}^{(j)}_{F} {\bm \Pi}) \Delta_j }{{2\omega_n^2}}+
 \frac{\hbar^2( {\bm v}^{(j)}_{F} {\bm \Pi}) ( {\bm v}^{(j)}_{F} {\bm \Pi}) \Delta_j}{4\omega_n^3}
 \end{eqnarray}
 and $ f^+_j ({\bm p},{\bm r}, \omega_n)  = f^*_j(-{\bm p},{\bm r}, \omega_n)$.
Then, for the summation over Matsubara frequencies, we get
\begin{equation}\label{}
 2\pi T \sum_{n=0}^{N_d} \omega_n^{-1} = G_0 + \tau
\end{equation}
where $\tau = (1- T/T_c)$.  We normalize gaps by $T_c/\sqrt{\rho}$, 
where
 \begin{equation}\label{Eq:rho}
 \rho =\sum_{n=0}^{\infty} \frac{\pi T_c^3}{\omega_n^{3}} = \frac{7\zeta (3)}{8\pi^2} \approx 0.1 ,
 \end{equation}
substitute
\Eqref{Eq:GLexpansion} into the self-consistency Eqs.\Eqref{Eq:SelfConsistentGap}
and get the system of GL equations
 \begin{equation}\label{Eq:GL3Band}
  \big[(G_0+\tau-\hat\Lambda^{-1}){\bm\Delta} \big]_j
  = -K^{(j)}_{ab}\Pi_a\Pi_b \Delta_j + |\Delta_j|^2\Delta_j  \,,
\end{equation}
Here ${\bm \Delta} = (\Delta_1, \Delta_2, \Delta_3)^T$, and the
anisotropy tensor is $K^{(j)}_{ab} =\hbar^2\rho \langle v^{(j)}_{Fa}
v^{(j)}_{Fb} \rangle /2T_c^2$ where the average is taken over the
$j$-th Fermi surface and $a,b$ stand for the $x,y$ coordinates.
The current is given by
 \begin{equation}\label{Eq:CurrentGL}
 {\bm j} ({\bm r})= \frac{4e\nu T_c^2}{\hbar\rho}
 \sum_i  \Im\; \Delta^*_i \hat K_i {\bm \Pi} \Delta_i .
 \end{equation}

%%%%%%%%%%%%%%%%%%%%%%%%%%%%%%%%%%%%%%%%%%%%

In the following, we consider coupling matrix  $\hat g= \nu_0 \hat{\Lambda}$ %$\hat{\Lambda}$
describing the case of an interband dominated pairing with
repulsion. We assume that the density of states is the same in all bands $\nu_{1,2,3}=\nu_0$ and parametrize
different pairing interactions with two dimensionless coefficients $\eta$ and $\nu$ as follows
\begin{equation}\label{Eq:Model3BandB1}
\hat\Lambda = - \left(%
\begin{array}{ccc}
  0        & \eta    & \lambda   \\
  \eta     & 0       & \lambda   \\
  \lambda  & \lambda & 0         \\
\end{array}  \right) \,.
\end{equation}
Here the coefficients describe different pairing channels in
$s_{\pm}+is$ and $s+id$ states. In the former case $\Delta_{1,2}$
correspond to the gaps at hole Fermi surfaces and $\Delta_3$ is
the gap at electron pockets so that  $u_{hh}=\nu_0\eta$ and
$u_{eh}=\nu_0\lambda$ are the hole-hole and electron-hole interactions
correspondingly. In contrast to describe the $s+id$ state we use
the same model \Eqref{Eq:Model3BandB1} but  assume that
$\Delta_{1,2}$ describe gaps in electron pockets and $\Delta_3$ is
the gap at the hole Fermi surface so that $u_{eh}=\nu_0\eta$ and
$u_{ee}=\nu_0\lambda$ to be electron-hole and electron-electron
interactions correspondingly.

Neglecting the r.h.s. in \Eqref{Eq:GL3Band} we get the linear
equation which determines the critical temperature
$G_0 =\min\left(G_1, G_2\right)$, where $G_1 = 1/\eta$ and
$G_2 =\left(\eta+\sqrt{\eta^2+8\lambda^2} \right) /4\lambda^2$
are the positive eigenvalues of the matrix
 \begin{equation}\label{Eq:Model3BandB2}
 \hat\Lambda^{-1} =  \frac{1}{2\lambda^2\eta}\left(
 \begin{array}{ccc}
  \lambda^2     & -\lambda^2    & -\lambda\eta  \\
  -\lambda^2    & \lambda^2     & -\lambda\eta  \\
  -\lambda\eta  & -\lambda\eta  & \eta^2        \\
\end{array}  \right) \,.
\end{equation}
The coupling matrix $\hat\Lambda^{-1}$ has only two positive
eigenvalues $G_{1,2}$ whose eigenvectors are
${\bm\Delta}_1=(-1,1,0)^T$ and ${\bm\Delta}_2=(x,x,1)^T$ with
$x=(\eta-\sqrt{\eta^2+8\lambda^2})/4\lambda$. Since only the
fields corresponding to the positive eigenvalues can nucleate,
the GL theory \Eqref{Eq:GL3Band} has to be reduced to a
two-component one.
To implement this reduction we represent the general order
parameter in terms of the superposition
\begin{equation}\label{Eq:OPreotation}
 {\bm \Delta} = \psi_1 {\bm \Delta}_1 + \psi_2 {\bm \Delta}_2\,,
\end{equation}
so that
\begin{equation}\label{Eq:fderSupp}
(\Delta_1,\Delta_2,\Delta_3)=(x\psi_2-\psi_1,x\psi_2+\psi_1,\psi_2)\,.
\end{equation}
where $\psi_{1,2}$ are the complex order parameter
fields that have different interpretation depending on the system
considered. In the $s+is$ case $\psi_1$ and $\psi_2$ are the order
parameter of $s_\pm$ pairing channels between two concentric hole
surfaces and between hole and electron surfaces correspondingly.
For the system with $s+id$ symmetry, $\psi_1$ is the order
parameter of the $d$ wave channel in electron pockets and $\psi_2$
is the order parameter of $s_{\pm}$ pairing between electron and
hole surfaces.

Now, substituting the ansatz \Eqref{Eq:OPreotation} into the system of 
Ginzburg-Landau equations \Eqref{Eq:GL3Band} we obtain, after projection 
on the vectors ${\bm \Delta}_{1,2}$, the system of two GL equations
\SubAlign{Eq:GL3BandReduced}{ 
 a_1\psi_1 + b_{1j}|\psi_j|^2\psi_1+b_{J}\psi_1^*\psi_2^2 &= \\
 ( K_{aa}^{(1)} + K_{aa}^{(2)} ) \Pi_a^2 \psi_1 
 &+ x ( K_{aa}^{(2)} - K_{aa}^{(1)} ) \Pi_a^2  \psi_2  \nonumber \\ 
   a_2\psi_2 + b_{2j}|\psi_j|^2\psi_2+b_{J}\psi_2^*\psi_1^2 &= \\
 \left[ x^2( K_{aa}^{(1)} + K_{aa}^{(2)} ) + K_{aa}^{(3)} \right] \Pi_a^2 \psi_2 
 &+ x ( K_{aa}^{(2)} - K_{aa}^{(1)} ) \Pi_a^2 \psi_1.  \nonumber
}
The parameters of the left hand side of the Ginzburg-Landau equations 
\Eqref{Eq:GL3BandReduced} are expressed, in terms of the coefficients 
of the coupling matrix \Eqref{Eq:Model3BandB1} as 
\SubAlign{Eq:Parameters}{
a_j &= -|\bm \Delta_j|^2(G_0-G_{j}+\tau )\,,~~ \\
&~~\text{with}~~ |\bm \Delta_1|^2  =2~~\text{and}~~ |\bm \Delta_2|^2=2x^2+1  \\
b_{11} &= 2    \,,~~~b_{22}=(2x^4+1) ~~\text{and}~~b_k:=b_{kk}\\
b_{12} &= 4x^2 \,,~b_J=2x^2 \,.
}
The system \Eqref{Eq:GL3BandReduced} is quite general and describes both 
$s+is$ and $s+id$ states. The difference between these two cases is determined 
by the symmetry of gradient terms. In $s+is$ state the $C_4$ symmetry requires 
that $K^{(j)}_{xx} = K^{(j)}_{yy} = K^{(j)}$. 
Then the general Eqs. \Eqref{Eq:GL3BandReduced} are
simplified as follows
 \begin{eqnarray}\label{Eq:GL3BandReducedSiS}
 a_1\psi_1 + b_{1j}|\psi_j|^2\psi_1 + b_{J}\psi_1^*\psi_2^2
 = k_{1j} {\bm \Pi}^2  \psi_j  \nonumber\\
 a_2\psi_2 + b_{2j}|\psi_j|^2\psi_2 + b_{J}\psi_2^*\psi_1^2
 = k_{2j} {\bm \Pi}^2 \psi_j
\end{eqnarray}
where
 \begin{eqnarray}\label{Eq:GradientCoeffitients}
  & k_{11} =    K^{(1)} + K^{(2)}  \nonumber \\
  & k_{22} =   [(K^{(1)} + K^{(2)})x^2 + K^{(3)} ] \nonumber \\
  & k_{12} =   x[ K^{(2)} - K^{(1)} ] .
 \end{eqnarray}

In $s+id$ state the $C_4$ symmetry is broken so that
$K^{(1,2)}_{xx} = K^{(2,1)}_{yy} = K^{(1,2)} $, where the bands $1,2$
correspond to the electron pockets. The hole band $3$ is considered to
be $C_4$ symmetric so that $K^{(3)}_{xx} = K^{(3)}_{yy} = K^{(3)}$.
Then we obtain from Eqs.\Eqref{Eq:GL3BandReduced}
  \SubAlign{Eq:GL3BandReducedSiD}{
   a_1\psi_1 + b_{1j}|\psi_j|^2\psi_1 + b_{J}\psi_1^*\psi_2^2&= \\ \nonumber
   k_{11} {\bm \Pi}^2  \psi_1 &+ k_{12} {\Pi}_{xy}  \psi_2 \\
   a_2\psi_2 + b_{2j}|\psi_j|^2\psi_2 + b_{J}\psi_2^*\psi_1^2&=\\ \nonumber
   k_{22} {\bm \Pi}^2  \psi_2 &+ k_{12} {\Pi}_{xy}  \psi_1
}
where the coefficients $k_{ab}$ are given by the same Eq.
\Eqref{Eq:GradientCoeffitients} as in $s+is$ state and
${\Pi}_{xy} = \Pi^2_x - \Pi^2_y $.

The general free energy functional which gives both the GL equations
(\ref{Eq:GL3BandReducedSiS},\ref{Eq:GL3BandReducedSiD}) is thus
given by  $F= \frac{\B^2}{8\pi} + B_0^2 \tilde{F}_s $ where   
 \begin{align}
 \tilde{F}_s &= \sum_{j=1}^2\Big\{  k_{jj}\left|\D\psi_j \right|^2
 +\alpha_j|\psi_j|^2+\frac{\beta_j}{2}|\psi_j|^4\Big\}   \nonumber\\
 &+k_{12,a}\Big((\Pi_a\psi_1)^*\Pi_a\psi_2+c.c. \Big)
 +\gamma|\psi_1|^2|\psi_2|^2                              \nonumber\\
 &+\frac{\delta}{2}\big(\psi_1^{*2}\psi_2^2+c.c.\big)   \,.\label{Eq:FreeEnergy}
 \end{align}
We denote $B_0 = T_c\sqrt{\nu_0/\rho}$  which is of the order the 
thermodynamical critical field at low temperatures \cite{Saint-James.Thomas.ea}. 
An extra factor of $\rho$ here comes from our normalization of gaps.  
 
In the case of $s+is$ symmetry, $k_{12,x}=k_{12,y}\equiv k_{12}$,
while for $s+id$ symmetry the mixed gradients satisfy
$k_{12,x}=-k_{12,y}\equiv k_{12}$. The other coefficients are given
by $\alpha_{k} = a_k $, $\beta_k= b_{kk}$, $\gamma = b_{12}$
and $\delta=b_J$.

\section{Time-dependent Ginzburg-Landau theory for multicomponent superconductors} 
\label{Appendix:TDGL}

The time-dependent GL theory formally applies only in the gapless regime when 
either the inelastic electron-phonon relaxation $\tau^{-1}_{ph}$ or spin flip 
$\tau^{-1}_{sf}$ rates are much larger than the superconducting pairing amplitude. 
We will consider here the former scenario which realizes in the vicinity of the
critical temperature $T_c-T\ll \tau^{-1}_{ph}$. In this case, the time-dependence 
is added to the generic Eqs.(\ref{Eq:GL3BandReducedSiS},\ref{Eq:GL3BandReducedSiD}) 
in quite a standard way, following the derivation for a single-component case 
\cite{KopninBook} 
\Align{Eq:GL3BandTD}{
 \big[(G_0+\tau-\hat\Lambda^{-1}){\bm\Delta} \big]_j &= \\ \nonumber
 \frac{\hbar\pi}{8T_c} \left(\frac{ \partial}{\partial t}+ \frac{2ie}{\hbar}\varphi \right) \Delta_j 
 &- K^{(j)}_{ab}\Pi_a\Pi_b \Delta_j + |\Delta_j|^2\Delta_j  \,,
}
Implementing the same reduction as in the stationary case, we obtain the system 
of two coupled TDGL equations:
\begin{equation}\label{Eq:AppTDGL}
 \Gamma_k \left(\frac{ \partial}{\partial t}+ \frac{2ie}{\hbar}\varphi \right)\psi_k
 =-\frac{\delta\F}{\delta\psi_k^*}
 \,,~~~ \frac{4\pi}{c} \Curl\B - \sigma_n \E={\bm j}_s
\end{equation}
where $k=1,\cdots,N$,  $\Gamma_k = |{\bm \Delta}_k|^2 \pi\hbar B_0^2/(8T_c) $ 
are the damping constants, $\varphi$ is the scalar potential of electric 
field, $\sigma_n$ is the normal state conductivity. The electric field is 
$\E=-c^{-1}\partial_t\A-\Grad \varphi$ where $\varphi$ is electrostatic 
potential and for the superfluid current we have an expression 
${\bm j}_s=-c\delta \F/\delta \A $. In the following since 
$|{\bm \Delta}_{1,2}|^2 \approx 1 $ we put $\Gamma_{1,2}=\Gamma= \pi\hbar B_0^2/(8T_c)$. 

In order to perform numerical simulations we normalize lengths by 
$ \tilde{\xi}_0 = \hbar \bar{v}_F/T_c$, where $\bar{v}_F$ is the average 
value of Fermi velocity, magnetic field by $B_0 = T_c\sqrt{\nu_0/\rho}$, 
free energy density by $F_0= B_{0}^2$, current density by
$j_0=c B_0/\xi_0$, time by $t_0 = \Gamma/B_0^2$, 
electric field by $E_0 = \xi_0 B_0/(t_0c)$, conductivity by 
$\sigma_{0}=c^2 \Gamma/(\xi_0^2B_0^2)$.
In such units the electron charge is replaced by an effective coupling 
constant $\tilde{e} = \pi B_0\xi_0^2/\Phi_0$ which parametrizes 
the regimes of extremely type-II and type-I superconductivity at $\tilde{e}\ll 1$ 
and $\tilde{e}\gg 1$ respectively.
To estimate the characteristic relaxation time we note that $t_0 = \pi\hbar/8T_c \sim 1 ps $ 
provided $T_c \sim 1$ meV which is about 10 K.  

The normal state electric conductivity $\sigma_n$ in \Eqref{Eq:AppTDGL} can be 
defined from the Drude model. Up to unimportant numerical coefficient resulting 
from the Fermi surface anisotropy, it yields $\sigma_n \sim e^2\nu_0 \bar{v}_F^2 \tau$, 
where $\tau$ is a transport time. This estimation does not depend on whether 
superconductor is in the clean or diffusive limit since $\sigma_n$ is a normal 
state characteristic. The dimensionless conductivity in our units is given by 
$\sigma_n/\sigma_0 \sim (\bar{v}_F/c)^2(\varepsilon_F/T_c)(\hbar\tau \varepsilon_F)$,
where $\varepsilon_F$ is the Fermi energy. Here the first factor is small since 
typically $(\bar{v}_F/c)^2 \sim 10^{-5}$ in metals. The last two factors are large 
since $(\varepsilon_F/T_c)\sim 10^2 $ and hence 
$(\hbar\tau \varepsilon_F) \sim 10^2 (\hbar\tau T_c)$. In this case in order to 
fulfil the clean limit conditions $(\tau T_c) >1$ we need to assume the values of 
dimensionless conductivity $\sigma_n/\sigma_0 > 0.1$.   

In dimensionless units the TDGL equations read
\begin{equation}\label{Eq:AppTDGL-dless}
  \left(\frac{ \partial}{\partial t}+ 2\tilde{e}\varphi \right)\psi_k=
  -\frac{\delta\F}{\delta\psi_k^*}
 \,,~~~  \Curl\B - \sigma_n \E={\bm j}_s
\end{equation}
where
$\E=-\partial_t\A-\Grad \varphi$ and ${\bm j}_s=
-\delta \F/\delta \A $.
The free energy is given by
\begin{align}
 \F &=\frac{\B^2}{8\pi}+\sum_{j=1}^2\Big\{  k_{jj}\left|\D\psi_j \right|^2
 +\alpha_j|\psi_j|^2+\frac{\beta_j}{2}|\psi_j|^4\Big\}
 \nonumber\\
 &+k_{12,a}\Big((\Pi_a\psi_1)^*\Pi_a\psi_2+c.c. \Big)
 +\gamma|\psi_1|^2|\psi_2|^2
 \nonumber\\
 &+\frac{\delta}{2}\big(\psi_1^{*2}\psi_2^2+c.c.\big)   \,.\label{Eq:FreeEnergy-dless}
 \end{align}
where ${\bm \Pi}=\Grad  -2 i \tilde{e} \A$.

%%%%%%%%%%%%%%%%%%%%%%%%%%%%%%%%%%%%%%%%%%%%%%%%%%%%%%%%%%%%%%%%%%%%%
%%%%%%%%%%%%%%%%%%%%%%%%%%%%%%%%%%%%%%%%%%%%%%%%%%%%%%%%%%%%%%%%%%%%%
\section{Numerical methods and starting guess}
\label{Appendix:Numerics}

We consider here the problem \Eqref{Eq:AppTDGL} defined
on a two-dimensional bounded domain $\Omega\subset\mathbbm{R}^2$
with $\partial\Omega$ its boundary. This problem is supplemented
by the insulator boundary conditions, that no supercurrent
flows through the boundary. The absence of supercurrent flowing
through the boundary reads as ${\bs n}\cdot\D\psi_a=0$ with
${\bs n}$ the normal vector to $\partial\Omega$.
The problem are then discretized using a finite element
formulation provided by the {\tt Freefem++} library \cite{Hecht:12}.
Discretization of the integration domain $\Omega$ is done using
a (homogeneous) triangulation over $\Omega$, based on
Delaunay-Voronoi algorithm. The fields are decomposed on
a continuous piecewise quadratic basis on each triangle.

\subsection*{Numerical procedure}

In our simulations, we chose the relaxation factor for the components 
of the order parameter $\Gamma=1$ and the normal state conductivity 
$\sigma_n=0.1$. We chose the gauge coupling constant $\tilde{e}=0.2$ 
and the coefficients of the coupling matrix \Eqref{Eq:Model3BandB1}, 
that determine the coefficients of the Ginzburg-Landau functional 
\Eqref{Eq:FreeEnergy-dless}, are $\eta=5$ and $\lambda=4.5$. The 
coefficients in GL functional \Eqref{Eq:FreeEnergy-dless} are 
then consistently obtained using the previously defined relations
and the obtained values are %$a_1=-1$, $a_2=-1.084$, 
$\beta_1=2$, $\beta_2=1.1079$, $\delta=0.4645$ and $\gamma=0.929$. 
Given the coefficients of gradient terms in different bands are 
$K^{(1)}=0.5$, $K^{(2)}=0.05$, $K^{(3)}=0.25$ and using the relation 
\Eqref{Eq:GL3BandReducedSiD}, the coefficients of the kinetic terms 
read as $k_1=0.55$, $k_2=0.375$ and $k_{12,x}=0.217$, while 
$k_{12,y}=k_{12,x}$ for $s+is$ states and $k_{12,y}=-k_{12,x}$ for $s+id$. 
We are interested here in the effect of inhomogeneities of the
order parameter induced by a local heating. In first approximation,
temperature dependence within the Ginzburg-Landau theory is
modelled by a linear dependence of the quadratic couplings on
temperature: $\alpha_k(T)=a_k-\tau$, where $\tau=\left(1- T/T_c \right)$
is the reduced temperature and $a_k$, a positive characteristic
constants determined from the microscopic calculations, see
Appendix \ref{Appendix:GL}). As the Ginzburg-Landau expansion
can be justified only in the vicinity of $T_c$, only a small
interval of $\tau$ can be investigated.

\begin{figure}[!htb]
\hbox to \linewidth{\hss
\includegraphics[width=0.7\linewidth]{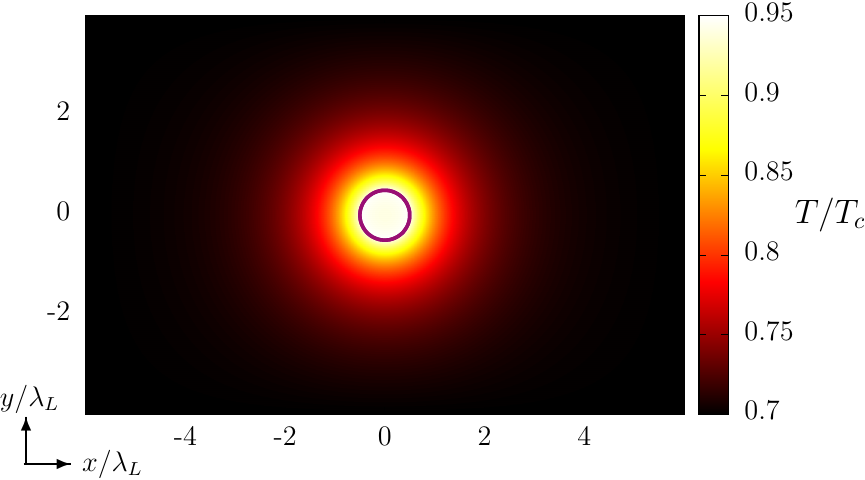}
\hss}
\caption{
(Color online) --
Temperature profile obtained by solving the stationary heat
equation \Eqref{Eq:Heat} for a local heating of the sample.
The inhomogeneous temperature profile is induced by the
ring-like heat source shown show by the (purple) circle.
}
\label{Fig:Temperature}
\end{figure}

The spatial modulation of the temperature is modelled as follow.
Let $\Gamma_0$ denote the outer boundary of the domain (the sample's
boundary) and $\Gamma_1$ an inner boundary (the heat source).
In our simulations we considered various shapes of $\Gamma_1$,
but the results reported here are for $\Gamma_1$ to be a small
circle. The temperature profile $T(\x)$ is determined by solving
the stationary heat equation with a small damping factor $\mu$
\Equation{Eq:Heat}{
\Delta T(\x)=\mu T(\x)~~~\text{and}~~~T(\Gamma_i)=\mathrm{cte}\,.
}
In our simulations, we set $T(\Gamma_0)=T_0=0.7T_c$ for the sample's
boundary. On the other hand, the temperature of the source $T_s$
varies with time as:
\Equation{Eq:AppTemperature}{
T_s(t)=T_0+\frac{T_1-T_0}{2}
\left(1-\cos\left(2\pi\frac{\lfloor{t}\rfloor}{n_T\Delta t}\right)\right)
}
where $\left \lfloor{x}\right \rfloor $ denotes the floor function,
$T_1=0.95T_c$ is the maximal temperature of the source and $n_T$ is
the number of heat source temperatures. In our simulations, we chose
$_T=30$. \Figref{Fig:Temperature} shows a typical solution for the
stationary heat equation \Eqref{Eq:Heat}. Note that our choice here
for considering the temperature profile given by the stationary heat
equation implies that we assume than heat transport occurs on
time scales much smaller than the other time scales of the
problem.

The heat equation \Eqref{Eq:Heat} is thus solved for a given
temperature $T_s$ of the heat source ($\Gamma_1$). This define
the temperature profile $T(\x,t)$ and thus the spatial
modulation quadratic couplings $\alpha_i(T)=a_i-\tau$.
The time dependent Ginzburg-Landau equations \Eqref{Eq:AppTDGL}
are discretized using Euler's forward method and iterated for
a given interval $\Delta t=N\delta t$. The time step $\delta t$
depends on the various time scales of the problem. Here we
chose $\delta t=10^{-2}$ and $N=8\times10^3$. After
$\Delta_t$, the procedure is repeated for a new temperature
$T_s$ of the heat source given by \Eqref{Eq:AppTemperature}.

\subsection*{Starting guess for domain-walls}

The starting solutions are stationary solutions of the GL
functional, for constant temperature. That is with no heat
source. They can simply be ground state but also topologically
non-trivial solutions in equilibrium. Domain-walls are topological
excitations that are associated with the spontaneous breakdown
of a discrete symmetry. These are field configurations that
interpolate between inequivalent ground-states that are
disconnected.
Here, we are in particular interested in domain-walls that
interpolate between regions with inequivalent relative
phases between the condensates. When the ground-state breaks
time-reversal symmetry, its complex conjugate is not a gauge
equivalent. That is, there exist no real number $\chi_0$ such
that $\Psi_0^*=\Exp{i\chi_0}\Psi_0$. If no such transformation
exist, then $\Psi_0^*\not\equiv\Psi_0$ and the configurations
with ground state phases $\bar{\varphi_a}$ and $-\bar{\varphi_a}$
are disconnected and degenerate in energy. Domain-walls that
interpolate between $\Psi_0$ and $\Psi_0^*$ are topologically
protected as their unwinding would require to overcome an
infinite energy barrier, see for example textbook discussion
in \cite{Manton.Sutcliffe,Rajaraman}. The domain-wall that
interpolates between $\Psi_0^*$ and $\Psi_0$ can thus be
parametrized by:
\Equation{Eq:PhaseDW}{
\psi_a^{(dw)}=u_a\exp\left[i\bar{\varphi_a}\tanh\left(
\frac{\x_\perp-\x_0}{\xi^{(dw)}_a} \right)\right]\,.
}
where $\x_0$ is the curvilinear abscissa that gives the
position of the domain-wall, and $\x_\perp$ is the
coordinate perpendicular to the domain-wall and $\xi^{(dw)}$
determines the width of the domain-wall.

\clearpage

\section{Description Additional Movie Material} 
\label{Appendix:Movie}

\begin{figure*}[b]
\hbox to \linewidth{\hss
\includegraphics[width=0.7\linewidth]{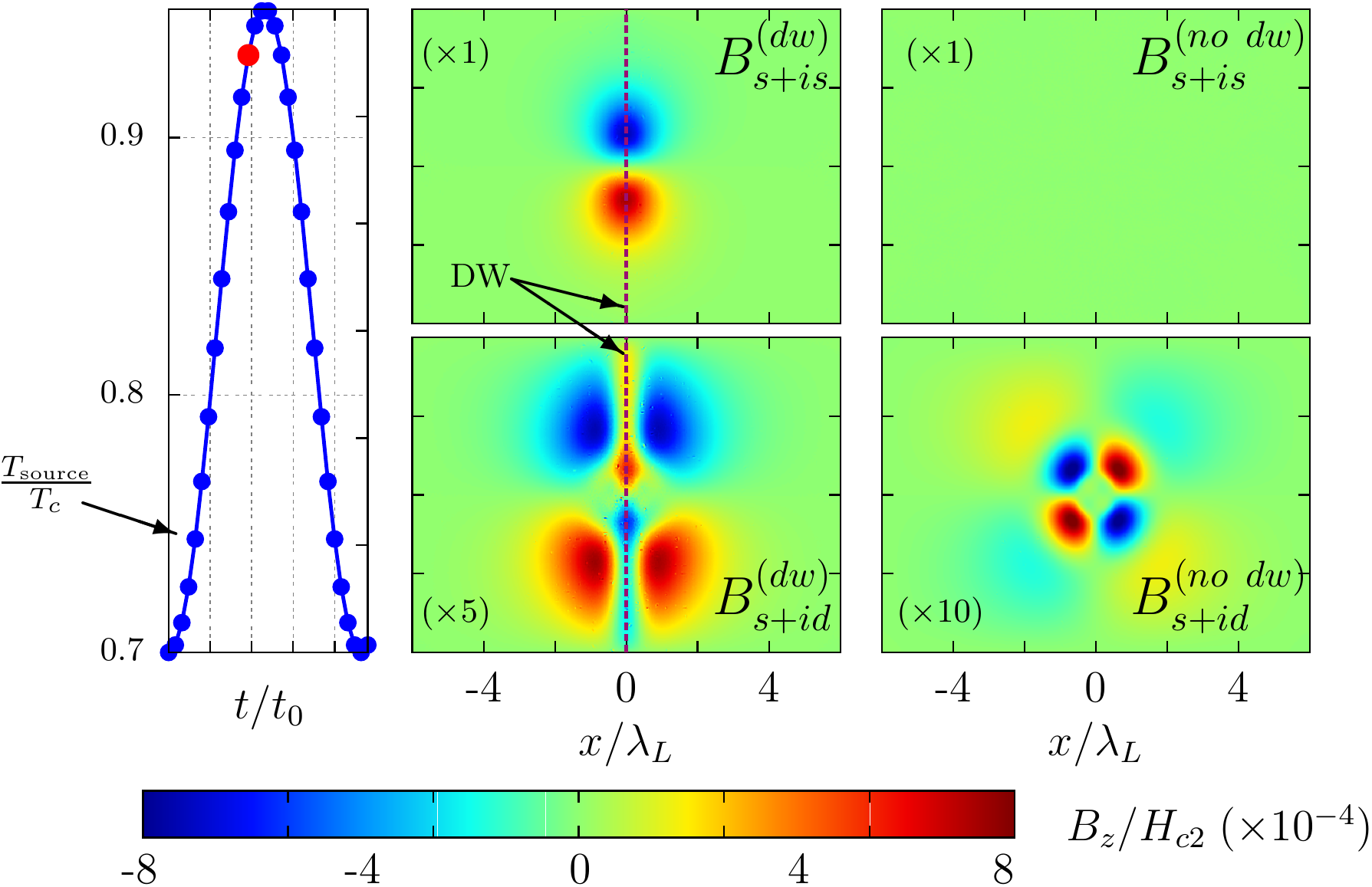}
\hss}
\caption{
(Color online) --
Evolution of the magnetic response that originates in time-varying local 
heating of the sample. The overall time sequence runs for $t/t_0=2400$.
The leftmost panel shows the time evolution of the source's temperature 
given by \Eqref{Eq:AppTemperature} and the red dot denotes the position 
in the time series. This corresponds to Figure 2 of the main text.
The color plot shows the magnitude of the out-of-plane induced magnetic 
field $B_z$ in unit of the second critical field. 
The two upper panels show the response of an $s+is$ superconducting state
respectively with a domain wall and the homogeneous state.
The two lower panels show the response of an $s+id$ superconducting state
respectively with a domain wall and the homogeneous (anisotropic) state.
}
\label{Fig:AnimMagnetic}
\end{figure*}

\begin{figure*}[!htb]
\hbox to \linewidth{\hss
\includegraphics[width=0.7\linewidth]{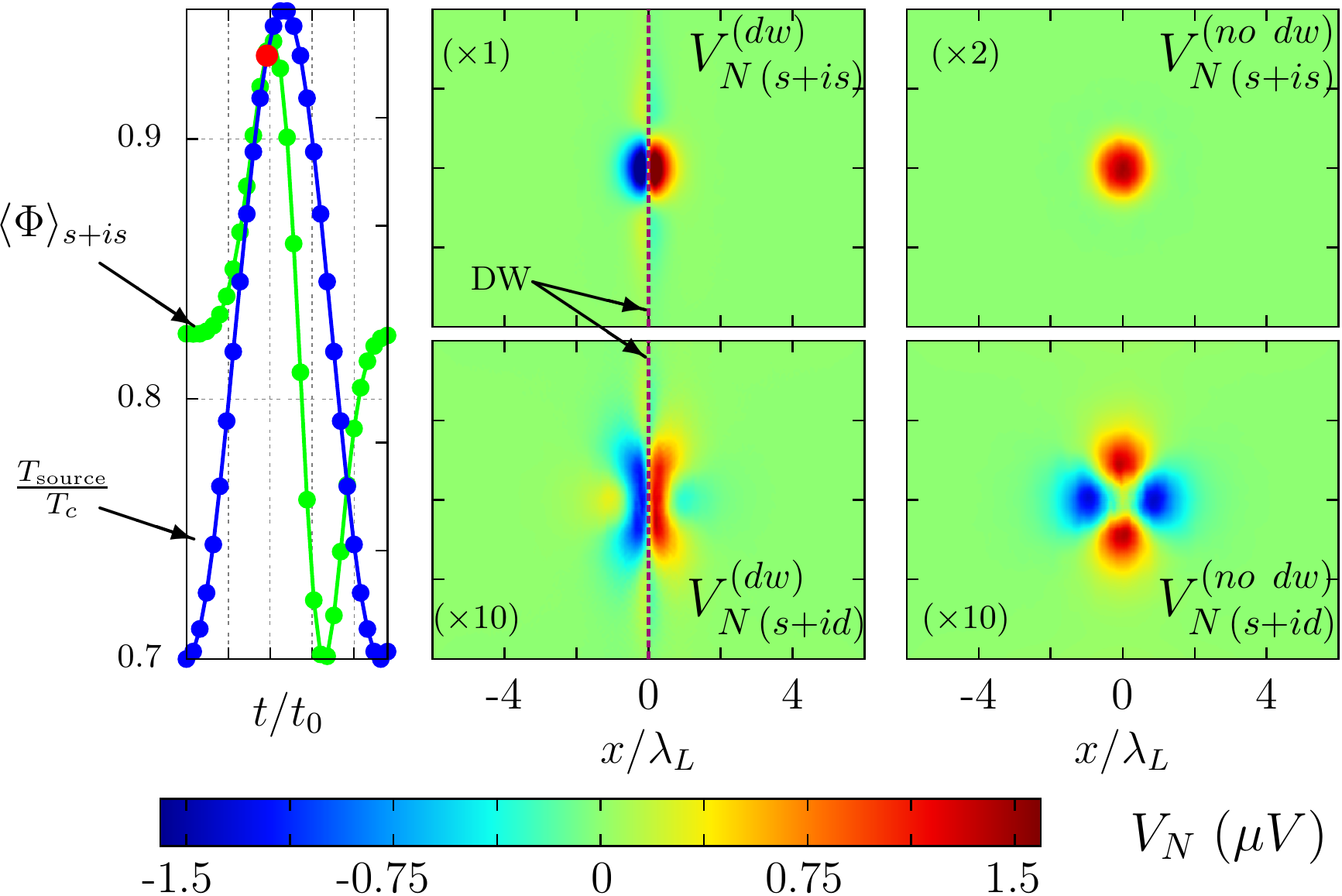}
\hss}
\caption{
(Color online) --
Evolution of the electric response that originates in time-varying local 
heating of the sample. The overall time sequence runs for $t/t_0=2400$.
The leftmost panel shows the time evolution of the source's temperature 
given by \Eqref{Eq:AppTemperature} and the red dot denotes the position 
in the time series. This corresponds to Figure 3 of the main text.
The color plot shows the voltage induced in the normal detector by the 
charge imbalance due to the non-stationary heating. 
The two upper panels show the response of an $s+is$ superconducting state
respectively with a domain wall and the homogeneous state.
The two lower panels show the response of an $s+id$ superconducting state
respectively with a domain wall and the homogeneous (anisotropic) state.
}
\label{Fig:AnimImbalance}
\end{figure*}

\begin{itemize}
\item Movie {\bf Anim-magnetic.avi}: \\
	shows the evolution of the magnetic response that originates 
	in time-varying local heating of the sample. Detailed description 
	is given in the caption of \Figref{Fig:AnimMagnetic}.
\item Movie {\bf Anim-imbalance.avi}: \\
	shows the evolution of the electric response that originates 
	in non-stationary local of the sample. Detailed description 
	is given in the caption of \Figref{Fig:AnimImbalance}.
\end{itemize}

%%%%%%%%%%%%%%%%%%%%%%%%%%%%%%%%%%%%%%%%%%%%%%%%%%%%%%%%%%%%%%%%%%%%%

\clearpage
%%%%%%%%%%%%%%%%%%%%%%%%%%%%%%%%%%%%%%%%%%%%%%%%%%%%%%%%%%%%%%%%%%%%%
%%%% Bibliography
%%\bibliographystyle{apsrev4-1}
%\bibliography{../../Thermophase}

\end{document}